\documentclass[]{IEEEphot}
\usepackage{algorithm}
\usepackage{booktabs}
\usepackage{amsfonts}
\usepackage{amsmath}
\usepackage{graphicx}
\usepackage{color}
\usepackage{multicol}
\usepackage{amssymb}
\usepackage{subfigure}
\usepackage{bm}
\usepackage{latexsym}
\usepackage{stfloats}
\usepackage{setspace}
\usepackage{cite}
\usepackage{changebar}
\jvol{}
\jnum{}
\jmonth{}
\pubyear{}

\begin{document}

\title{Performance of a Free Space Optical Relay-Assisted Hybrid RF/FSO System in Generalized M-Distributed Channels}

\author{Lei~Kong,$^{1}$~\IEEEmembership{Student~Member,~IEEE}, Wei~Xu,$^{1}$~\IEEEmembership{Senior~Member,~IEEE},
Lajos~Hanzo,$^{2}$~\IEEEmembership{Fellow,~IEEE}\\
Hua~Zhang,$^{1}$~\IEEEmembership{Member,~IEEE}, Chunming~Zhao,$^{1}$~\IEEEmembership{Member,~IEEE}}

\affil{$^{1}$National Mobile Communication Research Laboratory, Southeast University,\\
Nanjing 210096, China\\
$^{2}$ Southampton Wireless, School of ECS, University of Southampton, SO17 1BJ, UK}

\doiinfo{DOI: 10.1109/JPHOT.2009.XXXXXXX\\
1943-0655/\$25.00 \copyright 2009 IEEE}%

\maketitle

\markboth{IEEE Photonics Journal}{Performance of a Free Space Optical}

\begin{receivedinfo}%
%This work was supported by the National Basic Research Program of China (973) under 2013CB329204, the National Natural Science Foundation of China under 61223001.
\end{receivedinfo}

\begin{abstract}
This paper investigates the average symbol error rate (ASER) performance of a dual-hop hybrid relaying system relying on both radio frequency (RF) and free space optical (FSO) links. Specifically, the RF link is used for supporting mobile communication, while the FSO link is adopted as the backhaul of the cellular infrastructure. Considering non-line-of-sight (NLoS) RF transmissions and a generalized atmospheric turbulence (AT) channel, the associated statistical features constituted of both the exact and the asymptotic moment generating functions (MGF) are derived in closed form. They are then used for calculating the ASER of M-ary phase shift keying (PSK), differentially encoded non-coherent PSK (DPSK) and non-coherent frequency-shift keying (FSK). A range of additional asymptotic expressions are also derived for all the modulation schemes under high signal-to-noise ratios (SNR). It is observed from the asymptotic analysis that the ASERs of all the modulation schemes are dominated by the average SNR of the RF link in the hybrid relaying system using a fixed relay gain, while in the relaying system using a dynamic channel dependent relay gain, the ASERs of all the modulation schemes depend both on the average SNR and on the AT condition of the FSO path. We also find that the fixed-gain relaying strategy achieves twice the diversity order of the channel-dependent relaying strategy albeit at the cost of requiring a high power amplifier (PA) dynamic range at the relay node. Furthermore, by comparing the asymptotic ASERs, we calculate the SNR differences between the different modulation schemes in both the fixed-gain and the channel-dependent relaying system. Finally, simulation results are presented for confirming the accuracy of our expressions and observations.
\end{abstract}


\section{Introduction}
Free space optical (FSO) communication has attracted increasing research interests as a benefit of its high data rate, enhanced security achieved in the unlicensed optical electro-magnetic spectrum and its moderate deployment cost \cite{Kedar2004, Survey2014, Zhou2013}. It is suitable for a wide range of applications, such as enterprise/building connectivity, the back-haul of cellular systems, redundant backup links and disaster recovery \cite{Survey2014}. Despite the above-mentioned advantages, FSO systems remain vulnerable to atmospheric turbulence (AT) due to the random refractive index variation caused by the inhomogeneities in the temperature and pressure of the atmosphere \cite{Andrews1998}.

A number of studies have characterized the effects of AT on the attainable FSO system performance, which adopted sophisticated techniques for mitigating the performance degradation imposed by AT \cite{Nistazakis2009, Li2013, Zhu2002, Bayaki2009}. In \cite{Nistazakis2009}, the average capacity of FSO systems was investigated under both weak and strong turbulence conditions, while in \cite{Li2013}, both the bit error rate (BER) performance as well as the channel capacity were analyzed for optical code-division multiple-access (OCDMA) systems. Spatial diversity reception was investigated in \cite{Zhu2002} with the aid of both maximum-likelihood detection and maximum-likelihood sequence detection for overcoming the turbulence-induced fading. In \cite{Bayaki2009}, the BER performance was characterized as a function of the diversity gain in a multiple-input multiple-output (MIMO) FSO system. Relaying techniques have been widely used for exploiting the resultant diversity gain and for mitigating channel fading in traditional radio frequency (RF) systems \cite{Suraweera2012, Xu2010}. By contrast, in FSO systems, relaying has also been involved for mitigating the AT-induced fading \cite{Relay_FSO, Tang2014}. Specifically, both the outage probability (OP) and the average BER of a dual-hop fixed-gain wireless relaying system were investigated in \cite{Suraweera2012}. The authors of \cite{Xu2010} analyzed the asymptotic achievable rate of relay selection strategies in amplify-and-forward (AF) MIMO two-hop networks relying on feedback. The achievable performance of a relay-aided FSO system was first studied in \cite{Relay_FSO} under weak turbulence conditions, while in \cite{Tang2014}, both the OP and the average symbol error rate (ASER) were investigated for a multi-hop FSO system relying on differential phase-shift keying (DPSK) under strong AT.

Recently, FSO links have been shown to achieve a promising performance in backhaul transmission between each infrastructure element in cellular networks, because the cost of deploying optical fiber is sometimes prohibitively high, especially in ultra-dense networks or in historic medieval cities\cite{Tang2014, Yang2014}. Therefore, a hybrid relay-aided system based on RF combined with FSO can be established for exploiting the complementary advantages of both RF and FSO systems \cite{Lee2011, Zedini2015, Zhang2015, Miridakis2014}. A dual-hop hybrid Rayleigh RF and FSO relaying system was first considered in \cite{Lee2011}, where the attainable outage performance was quantified. Recently, this was further extended in \cite{Zedini2015} to a generalized Nakagami-m channel for the RF link, where both the outage and the ASER of binary modulation schemes were analyzed. Then a similar extension to the $\kappa-\mu$ or $\eta-\mu$ distributed RF channel was provided in \cite{Zhang2015}. In \cite{Miridakis2014}, analytical expressions of both the end-to-end outage and the ASER of binary phase-shift keying (BPSK) were derived for a multiuser hybrid RF/FSO relaying system. However, all these papers considered only simple binary modulation schemes, and assume the Gamma-Gamma (G-G) distributed FSO channel, which is suitable for modeling moderate to strong AT conditions.

In a subcarrier intensity modulation (SIM) based FSO application, M-ary phase-shift keying (MPSK), DPSK and non-coherent frequency-shift keying (NCFSK) have been advocated as their benefits like high efficiency, non-sensitive to phase ambiguous and/or lower implementation complexity \cite{Huang1993, Song2012}. Against this background, our contributions are as follows: we study the ASER performance of various modulation schemes in a dual-hop hybrid RF/FSO relaying system relying on both fixed-gain and on channel-dependent schemes. In particular, we consider a generalized AT model, namely the M-distribution, which has recently been presented in \cite{Navas2011, Navas_BER_2011} as a benefit of its excellent matching to experimental data, and because it is capable of characterizing most of the existing AT models, including the G-G and the K distribution. Under this scenario, we derive both the exact and approximate moment generating functions (MGF) of the end-to-end instantaneous signal-to-noise ratios (SNR) of both fixed-gain and channel-dependent hybrid relaying system. These MGFs are then used for deriving both the exact expression and high accuracy ASER approximations for MPSK. By using the end-to-end cumulative distribution function (CDF), we also deduce a closed-form ASER expression for DPSK/NCFSK. Moreover, the asymptotic ASERs of both MPSK and DPSK/NCFSK are studied with the aid of insightful observations at high SNRs for both relaying systems. We observe that the ASERs of all the modulation schemes are dominated by the average SNR of the RF link in the fixed-gain relaying system, whereas the ASERs of all the modulation schemes are related to the average SNR and the AT conditions of the FSO hop in the channel-dependent relaying system. Furthermore, we evaluate the SNR differences of different modulation schemes from the asymptotic expressions, which is helpful for system analysis and design.

The rest of the paper is organized as follows. Section II describes the system model of the hybrid RF/FSO relaying system. A range of statistical characteristics and both exact as well as asymptotic ASERs of the different modulation schemes are analyzed for the fixed-gain hybrid relaying system in Section III. In Section IV, the ASERs of different modulation schemes are studied for the channel-dependent hybrid relaying system. Our simulation results are presented in Section V, while Section VI concludes the paper. A list of all variables and functions used in this paper is presented in Table I.
\begin{table}\centering
\caption{Notations}\label{Tab_ED_4_ary}
\begin{tabular}{|l|l|}
\hline
 $\gamma$ & instantaneous ene-to-end SNR \\
 $\gamma_1$ & instantaneous SNR of the RF link \\
 $\gamma_2$ & instantaneous SNR of the FSO link \\
 $\Gamma_1$ & average SNR of the RF link \\
 $\Gamma_2$ & average SNR of the FSO link \\
 $G$ & fixed relaying gain \\
 $I$ & irradiance intensity \\
 $U_L$ & LoS component in the M-distribution \\
 $U_S^C$ & scattering component coupled to the LoS component in the M-distribution \\
 $U_S^G$ & scattering component independent of the LoS component in the M-distribution \\
 $\Omega$ & average optical power of the LoS component \\
 $2b_0$ & average optical power of the scatter component \\
 $\Omega'$ & average optical power of the LoS component and the scattering one coupled with it \\
 $\Phi_A$ & phase of the LoS component \\
 $\Phi_B$ & phase of the scattering component coupled with the LoS component \\
 $A$ & parameter of the M-distribution \\
 $\alpha$ & effective number of large scale cells\\
 $\beta$ & natural number, represents the fading parameter \\
 $\xi$ & average power of classic scattering component \\
 $\rho$ & ratio of the scattering power coupled with the LoS component \\
 $\eta$ & electrical-to-optical conversion efficiency \\
 $\sigma_i^2$ &  variance of the additive white Gaussian noise, $i=1$ for RF link, $i=2$ for FSO link \\
 $\mathbb{E}[.]$ & expectation operator \\
 $f(.)$ & probability density function \\
 $F(.)$ & cumulative distribution function \\
 $M(.)$ & moment generating function \\
 $K_v(.)$ & modified Bessel function of the second kind with order $v$ \\
 $G(.)$ & Meijer's-G function\\
\hline
\end{tabular}
\end{table}

\section{System Model}
\begin{figure} \centering
  \includegraphics[scale = 0.6]{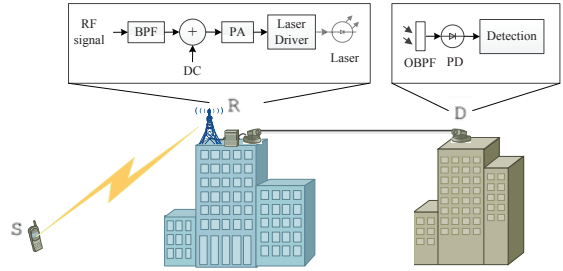}
  \caption{The dual-hop hybrid RF/FSO relaying system, including the block diagrams of the relay node and the destination node.}\label{fig_System_Model}
\end{figure}
We consider the dual-hop hybrid RF/FSO relaying system of Fig.~\ref{fig_System_Model}, where the source node (SN) \texttt{S} communicates with the destination node (DN) \texttt{D} via the intermediate relay node (RN) \texttt{R} by adopting an amplify-and-forward (AF) relaying scheme. Let $s$ be the normalized signal transmitted from \texttt{S} to \texttt{R} through an RF link. The signal received at \texttt{R} is expressed as:
\begin{equation}\label{eq_Rx_Relay}
    r_{\texttt{R}} = h s+ n_1,
\end{equation}
where $h$ represents the fading of the RF link, which is a complex Gaussian random variable \cite{Lee2011, Samimi2013}, and $n_1$ represents the additive white Gaussian noise (AWGN) with variance of $\sigma_1^2$ and zero mean.

The SIM scheme is employed at \texttt{R}, where a standard RF coherent/noncohenrent modulator and demodulator can be used for transmitting and recovering the source data \cite{Huang1993, Popoola2009}. At \texttt{R}, after filtering by a bandpass filter (BPF), a direct current (DC) bias is added to the filtered RF signal to ensure that the optical signal is non-negative. Then the biased signal is amplified and sent to a continuous wave laser driver. The output optical signal is written as \cite{Lee2011}:
\begin{equation}\label{eq_Tx_Relay}
    s_{\texttt{R}} = G (1+\eta r_{\texttt{R}}),
\end{equation}
where $G$ is the relay amplification gain at \texttt{R} and $\eta$ denotes the electrical-to-optical conversion efficiency. Then the optical signal is forwarded from \texttt{R} to \texttt{D} through an FSO channel.

After removing the DC component at \texttt{D}, the received signal is given by:
\begin{equation}\label{eq_Rx_Destination}
    r_{\texttt{D}} = I Gs_{\texttt{R}} + n_2 - IG = I G \eta ( h s+ n_1) + n_2,
\end{equation}
where $I$ is the irradiance intensity along the FSO link and $n_2$ represents the AWGN with a variance of $\sigma_2^2$ and zero mean in the FSO link.

The instantaneous end-to-end SNR of the relay-aided system is given by \cite{Peppas2013, Hasna2004}:
\begin{equation}\label{eq_SNR}
    \gamma = \frac{I^2 G^2 \eta^2 h^2 }{I^2G^2\eta^2 \sigma_1^2 + \sigma_2^2} = \frac{\frac{ h^2}{\sigma_1^2} \frac{\eta^2I^2}{\sigma_2^2}}{\frac{\eta^2I^2}{\sigma_2^2}+\frac{1}{G^2\sigma_1^2}}.
\end{equation}

Regarding the fixed-gain relay strategy at the RN, its relay gain is fixed to a constant value, which is independent of the channel state information (CSI) of the first-hop channel. We fix the relay amplification gain to $G^2 = \frac{1}{C \sigma_1^2}$, where $C$ is a constant parameter \cite{Hasna2004, Suraweera2009}. The end-to-end SNR of the relaying system is then rewritten as:
\begin{equation}\label{eq_SNR_Fixed}
    \gamma = \frac{\gamma_1\gamma_2}{\gamma_2+C},
\end{equation}
where we define $\gamma_1 = \frac{h^2}{\sigma_1^2}$ and $\gamma_2 = \frac{\eta^2I^2}{\sigma_2^2}$ representing the instantaneous SNR of the RF and FSO links, respectively. Since the RN amplification gain $G$ is fixed, the actual forwarded signal has a varying output power, because the relaying signals have been affected by the first hop channel fading before their fixed-gain amplification at the  RN. Hence, the RN has to have a power amplifier (PA) exhibiting a high dynamic range such as a linear class-A amplifier.

For the other scenario, where the CSI of the first hop is available, the channel-dependent relay strategy adjusts the relay gain according to the CSI of the RF link, thus resulting in a fixed output signal power at the RN. Accordingly, the relaying gain is given by $G^2 = \frac{1}{h^2+\sigma_1^2}$ \cite{Hasna2004}. By substituting it into \eqref{eq_SNR}, the end-to-end SNR of the channel-dependent relaying scheme becomes:
\begin{equation}\label{eq_SNR_CSI}
    \gamma =\frac{\gamma_1\gamma_2}{\gamma_1+\gamma_2+1}.
\end{equation}

\subsection{RF link}
In an urban environment, the RF transmission links spanning from \texttt{S} to \texttt{R} are subjected to multi-path fading in non-line-of-sight (NLoS) links, which can be characterized by the classic Rayleigh distribution. Accordingly, the instantaneous SNR of the RF link obeys an exponential distribution with its probability density function (PDF) given by:
\begin{equation}\label{eq_Rayleigh_PDF}
    f_{\gamma_1}(\gamma_1) = \frac{1}{\Gamma_1}\exp\left(-\frac{\gamma_1}{\Gamma_1}\right),
\end{equation}
where $\Gamma_1 = \frac{P_1}{\sigma_1^2}$ is the average SNR of the RF link.

\subsection{FSO link}
The FSO link is used as the \texttt{R} to \texttt{D} back-haul. Generally, the AT caused by the random refractive index variation of the atmosphere is considered as one of the most grave impairments of the FSO link. In \cite{Navas2011}, a generalized statistical model, namely the M-distribution, has been shown to exhibit an excellent fit to experimental propagation measurements, hence accurately characterizing the AT. In the M-distribution, the observation field of the irradiance intensity experienced at the receiver consists of three different components: the line-of-sight (LoS) component $U_L$, the scattering term $U_S^C$ coupled to the LoS component and the classic scattering term $U_S^G$ independent of the LoS component. The average optical power of the LoS term is given by $\Omega = \mathbb{E}[|U_L|^2]$, where $\mathbb{E}[.]$ is the expectation operator. The average optical power of all the scattering related terms is denoted by $2b_0 = \mathbb{E}[|U_S^C|^2+|U_S^G|^2]$. The PDF of the normalized irradiance $I$ of the M-distribution is expressed as \cite{Navas2011, Navas_BER_2011}:
\begin{equation}\label{eq_M_PDF}
    f_{I}(I) = A \sum_{k=1}^\beta a_k I^{\frac{\alpha+k}{2}-1}K_{\alpha-k}\left(2\sqrt{\frac{\alpha\beta I}{\xi \beta+\Omega'}}\right),
\end{equation}
where we have
\begin{equation}\label{eq_M_PDF_Condition}
    \begin{cases}
    A = \frac{2\alpha^{\alpha/2}}{\xi^{1+\alpha/2}\Gamma(\alpha)}\left(\frac{\xi\beta}{\xi \beta+\Omega'}\right)^{\frac{\alpha}{2}+\beta}, \\
    a_k = \left( _{k-1 }^{\beta-1}\right)\frac{\left(\xi\beta+\Omega'\right)^{1-\frac{k}{2}}}{(k-1)!} \left(\frac{\Omega'}{\xi}\right)^{k-1}\left(\frac{\alpha}{\beta}\right)^{\frac{k}{2}},
    \end{cases}
\end{equation}
with $\alpha$ being a positive parameter related to the effective number of large-scale cells of the scattering process, $\beta$ is a natural number representing the fading parameter, $\xi=2b_0(1-\rho)$ represents the average power of the classic scattering component $U_S^G$, $\rho$ is the ratio of the power of the scattering component coupled with LoS to that of all scattering components, and $\Omega' = \Omega + \rho2b_0 + 2\sqrt{\rho2b_0\Omega}\cos(\Phi_A-\Phi_B)$ denotes the average optical power of the coherent contributions, including both the LoS component and the scattering component coupled with it, where $\Phi_A$ and $\Phi_B$ are, respectively, the deterministic phase of the LoS component and the scattering component coupled with it \cite{Navas2011}. Finally, $K_v(.)$ is the modified Bessel function of the second kind with order $v$.

Note that a generalized expression of the M-distribution has been provided in \cite[Eq. (22)]{Navas2011}. Here, the expression in \eqref{eq_M_PDF} is a particularization of the generalized expression, where $\beta$ is confined to an integer value. Based on \cite{Navas_BER_2011}, this particularization is capable of representing every turbulence scenario of the generalized model, while enjoying the advantage of avoiding infinite summation. Specifically, some other popular distribution models of the AT, such as the Gamma-Gamma distribution and K distribution, can be treated as special cases of the M-distribution. Table.~\ref{Tab_Special_M} summarizes the special cases.

\begin{table}\centering
\caption{Special cases of M-distribution}\label{Tab_Special_M}
\begin{tabular}{l|c}
\hline \toprule
  Distribution & Conditions   \\
\hline
  Gamma-Gamma & $\rho = 1$, $\Omega' = 1$ \\
  K           & $\rho = 0$, $\Omega = 0$ \\
\hline \toprule
\end{tabular}
\end{table}

The PDF of the instantaneous electronic SNR associated with the FSO link is readily obtained with the aid of \eqref{eq_M_PDF} as follows \cite{Nistazakis2009}:
\begin{equation}\label{eq_M_SNR_PDF}
    f_{\gamma_2}(\gamma_2) = \frac{A}{2} \sum_{k=1}^\beta a_k \frac{\gamma_2^{\frac{\alpha+k}{4}-1}}{\Gamma_2^{\frac{\alpha+k}{4}}}K_{\alpha-k}\left(2\sqrt{\Lambda\sqrt{\frac{\gamma_2}{\Gamma_2}}}\right),
\end{equation}
where we have $\Lambda = \frac{\alpha\beta}{\xi\beta+\Omega'}$ and $\Gamma_2 =  \eta^2 \mathbb{E}[I]^2/\sigma_2^2 = \eta^2/\sigma_2^2$ is the electronic average SNR in the FSO link.

\section{Performance of the Hybrid RF/FSO System with Fixed-Gain Relaying}
Under the idealized simplifying assumption of having perfect channel estimates, the family of modulation schemes relying on coherent detection tends to require a lower SNR than their non-coherently detected counterparts, such as MPSK. However, the phase recovery error of coherent detection degrades the attainable system performance, while differential detection of DPSK is less susceptible to it. In practice, the uncertainty of the carrier synchronization and the carrier recovery error will make non-coherent modulation, such as NCFSK, a better choice. Moreover, the non-coherent detection also reduces the complexity of the receiver. Hence, non-coherent schemes constitute an attractive design alternative in some FSO applications \cite{Song2012}. In this section, the ASERs of MPSK, DPSK and NCFSK are investigated analytically in our fixed-gain hybrid RF/FSO relaying system.

\subsection{ASER Performance of MPSK}
To analyze the ASER performance of MPSK in the fixed-gain relaying system, we commence by deriving some useful statistical features of the end-to-end SNR.

\subsubsection{Statistical Characteristics}
The CDF of the end-to-end SNR $\gamma$ for the fixed-gain relaying system has been derived in \cite{Samimi2013}:
\begin{align}\label{eq_gamma_CDF}
    F_{\gamma}(\gamma)&= 1 - \frac{A\exp(-\frac{\gamma}{\Gamma_1})}{8\pi}\sum_{k=1}^{\beta}a_k \left(\frac{C\gamma}{\Gamma_1 \Gamma_2}\right)^{\frac{\alpha+k}{4}}  G_{0,5}^{5,0}\left(\Lambda^2\frac{C\gamma}{16\Gamma_1\Gamma_2} |_{\kappa_1}^{-}\right),
%     \overset{(a)}{=} 1 - \frac{A\exp(-\frac{\gamma}{\Gamma_1})}{8\pi}\sum_{k=1}^{\beta}a_k 2^{\alpha+k}\Lambda^{-\frac{\alpha+k}{2}}G_{0,5}^{5,0}\left(\Lambda^2\frac{C\gamma}{16\Gamma_1\Gamma_2} |_{\kappa_2}^{-}\right),
\end{align}
where $G\left(x\right)$ is Meijer's G-function and $\kappa_1=\frac{\alpha-k}{4},\frac{\alpha-k+2}{4},\frac{k-\alpha}{4},\frac{k-\alpha+2}{4},\frac{-\alpha-k}{4}$.  For simplicity, by using \cite[Eq. (07.34.17.0011.01)]{Wolfram}, the above CDF can be rewritten as
\begin{align}\label{eq_gamma_CDF}
    F_{\gamma}(\gamma) &=  1 - \frac{A\exp(-\frac{\gamma}{\Gamma_1})}{8\pi}\sum_{k=1}^{\beta}a_k 2^{\alpha+k}\Lambda^{-\frac{\alpha+k}{2}}  G_{0,5}^{5,0}\left(\Lambda^2\frac{C\gamma}{16\Gamma_1\Gamma_2} |_{\kappa_2}^{-}\right),
\end{align}
where $\kappa_2 = \frac{\alpha}{2},\frac{\alpha+1}{2}, \frac{k}{2}, \frac{k+1}{2}, 0$.

To derive the ASER for different modulation schemes, it is beneficial to use the moment generating function (MGF) of the end-to-end SNR $\gamma$. By using \eqref{eq_gamma_CDF}, the MGF is derived as follows:
\begin{align}\label{eq_MGF_Re}
    M_{\gamma}(s) &= \mathbb{E}[e^{-\gamma s}] \overset{(a)}{=} s\int_{0}^{\infty}e^{-s\gamma}F_{\gamma}(\gamma){d}\gamma \nonumber \\
    &= s \int_0^{\infty} \exp(-\gamma s)d{\gamma} - \frac{sA}{8\pi}\sum_{k=1}^{\beta}a_k 2^{\alpha+k}\Lambda^{-\frac{\alpha+k}{2}} \int_{0}^{\infty}  \exp(-s\gamma-\frac{\gamma}{\Gamma_1}) G_{0,5}^{5,0}\left(\frac{\Lambda^2 C\gamma}{16\Gamma_1\Gamma_2} |_{\kappa_1}^{-}\right){d}\gamma \nonumber \\
    & \overset{(b)}{=} 1 - \frac{2^{\alpha}A\Gamma_1s}{8\pi (1+\Gamma_1s)}\sum_{k=1}^{\beta}a_k2^{k}\Lambda^{-\frac{\alpha+k}{2}} G_{1,5}^{5,1}\left(\frac{\Lambda^2 C}{16 \Gamma_2(1+\Gamma_1s)} |_{\kappa_2}^{0}\right),
\end{align}
where Equation $(a)$ is obtained using integration by parts and $(b)$ is calculated by using the integral \cite[Eq. (7.813-1)]{Integral_Table}.

\subsubsection{ASER Analysis} \label{Sec_ASER_MPSK}
The ASER of MPSK can be expressed in terms of the MGF, which is given by \cite{Simon1995}:
\begin{equation}\label{eq_SEP_MPSK}
    P_s^{MPSK} = \frac{1}{\pi}\int_0^{\Theta}M_{\gamma}\left(\frac{n^2}{\sin^2\theta}\right){d}\theta,
\end{equation}
where $\Theta=\frac{(M-1)\pi}{M}$ and $n = \sin(\pi/M)$. By substituting \eqref{eq_MGF_Re} into \eqref{eq_SEP_MPSK}, we arrive at the ASER of MPSK in the form of the integral expression of:
\begin{align}\label{eq_SEP_MPSK_Re}
P_s^{MPSK} &= \frac{1}{\pi}\int_0^{\Theta}1 {d}\theta - \frac{2^{\alpha}A\Gamma_1n^2}{8\pi^2}\sum_{k=1}^{\beta}a_k2^{k} \Lambda^{-\frac{\alpha+k}{2}} \int_0^{\Theta}\frac{1}{\Phi} G_{1,5}^{5,1}\left(\frac{\Lambda^2 C \sin^2\theta}{16 \Gamma_2\Phi} |_{\kappa_2}^{0}\right){d}\theta,
\end{align}
where we have $\Phi = \sin^2\theta+\Gamma_1n^2$. The integral above cannot be readily expressed in a closed-form expression. Fortunately, from \cite{Mckay2009}, an accurate ASER approximation can be derived for MPSK by using the MGF:
\begin{align}\label{eq_SER_MPSK_Approx}
   \hat{P_s}^{MPSK} &= \left(\frac{M-1}{2M}-\frac{1}{6}\right)M_{\gamma}(s_1)+\frac{1}{4}M_{\gamma}\left(s_2\right) + \left(\frac{M-1}{2M}-\frac{1}{4}\right) M_{\gamma}\left(s_3\right),
\end{align}
where we have $s_1 = \sin^2(\pi/M), s_2 = \frac{4\sin^2(\pi/M)}{3} = \frac{4s_1}{3}$, and $s_3 = \frac{\sin^2(\pi/M)}{\sin^2(\pi(M-1)/M)} = 1$. Substituting  \eqref{eq_MGF_Re} into \eqref{eq_SER_MPSK_Approx}, the approximate ASER of MPSK can be readily calculated. As it will be shown in Section V, the ASER approximation in \eqref{eq_SER_MPSK_Approx} perfectly matches the exact ASER of MPSK cross the entire SNR range.

Note that the M-distribution is a general model which is capable of characterizing most of the popular distributions including the K-distribution and the G-G distribution. Therefore, the ASER performance of MPSK under both the K and G-G distributions can be deduced by our derived results. In particular, when we only encounter the classic scattering component in the FSO links, i.e. we have $\rho = 0$ and $\Omega = 0$ in \eqref{eq_M_PDF} and \eqref{eq_M_PDF_Condition}, the M-distribution reduces to the K-distribution. In this situation, $Aa_k\Lambda^{-\frac{\alpha+k}{2}}$ in \eqref{eq_MGF_Re} equals to zero, except for $k = 1$. After some further manipulations, we arrive at $Aa_1\Lambda^{-\frac{\alpha+1}{2}} = \frac{2}{\Gamma(\alpha)}$. Therefore, the MGF in \eqref{eq_MGF_Re} reduces to
\begin{align}\label{eq_MGF_K}
    M_{\gamma}^{K}(s) &= 1 - \frac{2^{\alpha}\Gamma_1s}{2\pi\Gamma(\alpha)(1+\Gamma_1s)} G_{1,5}^{5,1}\left(\frac{\alpha^2 C}{64 b_0^2  \Gamma_2(1+\Gamma_1s)} |_{\kappa_{K}}^{0}\right),
\end{align}
where $\kappa_{K} = \frac{\alpha}{2}, \frac{\alpha+1}{2}, \frac{1}{2}, 1, 0$. Now, both the exact and the approximate ASER of MPSK can be obtained for the K-distributed FSO channel by substituting \eqref{eq_MGF_K} into \eqref{eq_SEP_MPSK} and \eqref{eq_SER_MPSK_Approx}, respectively.

Considering another case when both the LoS component and the scattering component coupled with the LoS affect the FSO link, the M-distribution retrieves to the G-G distribution with $\rho=1$ and $\Omega'=1$. Under this scenario, $Aa_k$ equals to zero, except for $k=\beta$. After some further manipulations, we arrive at $Aa_\beta = 2(\alpha\beta)^{(\alpha+\beta)/2}/[\Gamma(\alpha)\Gamma(\beta)]$. Therefore, the MGF in \eqref{eq_MGF_Re} reduces to
\begin{align}\label{eq_MGF_GG}
    M_{\gamma}^{GG}(s) &= 1 - \frac{2^{\alpha+\beta}\Gamma_1s}{4\pi\Gamma(\alpha)\Gamma(\beta)(1+\Gamma_1s)} G_{1,5}^{5,1}\left(\frac{\Lambda^2 C}{16  \Gamma_2(1+\Gamma_1s)} |_{\kappa_{GG}}^{0}\right),
\end{align}
where we have $\kappa_{GG} = \frac{\alpha}{2}, \frac{\alpha+1}{2}, \frac{\beta}{2}, \frac{\beta+1}{2}, 0$. Accordingly, both the exact and the approximate SER of MPSK can be obtained for transmission over the G-G channel by substituting \eqref{eq_MGF_GG} into \eqref{eq_SEP_MPSK} and \eqref{eq_SER_MPSK_Approx}, respectively.

\subsubsection{Asymptotic Characterization}
The above ASER expression of MPSK is derived in the form of Meijer's G-function. Although Meijer's G-function can be expressed in terms of more popular hypergeometric functions, it appears hard to gain deeper insights. We circumvent the problem by investigating the asymptotic performance of MPSK at high SNRs. Generally, it is challenging to characterize the asymptotic performance of MPSK by directly evaluating the ASER in \eqref{eq_SEP_MPSK_Re} and \eqref{eq_SER_MPSK_Approx}. Instead, we resort to characterizing the asymptotic approximation of the MGF at high SNRs, which is shown to be effective for obtaining the asymptotic ASER.

%To facilitate the asymptotic analysis, we assume $\Gamma = \Gamma_1 = \lambda \Gamma_2$ where $\lambda$ is a constant indicating the ratio of the average SNRs of the two hop channels.
By observing the MGF expression in \eqref{eq_MGF_Re}, we find that at high SNR the Meijer's G-function can be expressed in form of its series representation \cite[Eq.~07.34.06.0006.01]{Wolfram} for $\Gamma_1, \Gamma_2\rightarrow\infty$. Considering the fact that $\alpha$ is larger than $\beta$ in the M-distribution, the series expansion of Meijer's G-function is dominated by the smallest parameter in $\kappa_2$, which yields:
\begin{align}\label{eq_MG_HighSNR}
&\quad\lim_{\Gamma_1,\Gamma_2\rightarrow\infty} G_{1,5}^{5,1}\left(\frac{\Lambda^2 C}{16 \Gamma_2(1+\Gamma_1 s)} |_{\kappa_2}^{0}\right)\nonumber \\
& =\Gamma\left(\frac{\alpha}{2}\right) \Gamma\left(\frac{\alpha+1}{2}\right) \Gamma\left(\frac{k}{2}\right) \Gamma\left(\frac{k+1}{2}\right) \nonumber \\
    & \overset{(c)}{=} 2^{2-\alpha-k}\pi\Gamma(\alpha)\Gamma(k),
\end{align}
where Equation $(c)$ is obtained by exploiting that $\Gamma(2x) = \frac{2^{2x-1}}{\sqrt{\pi}}\Gamma(x)\Gamma(x+1/2)$.
By substituting \eqref{eq_MG_HighSNR} and \eqref{eq_M_PDF_Condition} into \eqref{eq_MGF_Re}, the asymptotic MGF becomes:
\begin{align}\label{eq_MGF_HighSNR}
    \tilde{M}_{\gamma}(s)&=\lim_{\Gamma_1,\Gamma_2\rightarrow\infty}M_{\gamma}(s) \nonumber \\
    &= 1 - \frac{\Gamma_1 s}{1+\Gamma_1 s}\sum_{k=1}^{\beta}\left( _{k-1 }^{\beta-1}\right)\frac{(\xi\beta)^{\beta-k}\Omega'^{k-1}}{ (\xi\beta+\Omega')^{\beta-1}} \nonumber \\
    &\overset{(d)}{=} 1 - \frac{\Gamma_1 s}{1+\Gamma_1 s} \overset{(e)}{=} \frac{1}{\Gamma_1 s} + O\left(\Gamma_1^{-2}\right),
\end{align}
where Equation $(d)$ follows from the binomial expression $\sum_{k=1}^{\beta}\left( _{k-1 }^{\beta-1}\right)(\xi\beta)^{\beta-k}\Omega'^{k-1} = (\xi\beta+\Omega')^{\beta-1}$, and Equation $(e)$ is obtained by retaining only the most dominant term for a high SNR $\Gamma_1$.

Now, we find from \eqref{eq_MGF_HighSNR} that the end-to-end MGF does not depend on the turbulence condition of the FSO link at high SNRs. By substituting \eqref{eq_MGF_HighSNR} into \eqref{eq_SEP_MPSK}, we arrive at the asymptotic ASER of MPSK:
\begin{align}\label{eq_SER_MPSK_Upper_Bound}
   \tilde{P}_{s}^{MPSK} &= \frac{1}{\Gamma_1 \pi} \int_{0}^{\Theta}\frac{\sin^2\theta}{n^2} {d}\theta = \frac{\Xi}{\Gamma_1},
\end{align}
where $\Xi = \frac{2 \Theta - \sin2\Theta}{4n^2 \pi}$ is a constant, which is dependent on the modulation order $M$.

%\begin{align}\label{eq_SER_MPSK_Upper_Bound}
%   \tilde{P}_{s}^{MPSK} &= \left(\frac{M-1}{2M}-\frac{1}{6}\right)\frac{1}{\Gamma s_1} +\frac{1}{4\Gamma s_2} \nonumber \\ &\quad +\left(\frac{M-1}{2M}-\frac{1}{4}\right)\frac{1}{\Gamma s_3}\nonumber \\ &= \frac{\Xi}{\Gamma},
%\end{align}
%where we have $\Xi = \frac{25M+12Ms_1-24s_1-24}{48Ms_1}$ is a constant dependent on the modulation order $M$.

\emph{Remark 1:} It can be observed from the above expression that the asymptotic ASER of MPSK is no longer dependent on the FSO turbulence, but it is dominated by the average SNR of the RF link. At high SNRs, the diversity order is an important measure of the SER performance, which is defined as the slope of the log-log SER versus average SNR value \cite{Narasimhan2006, Xia2011}. We arrive at the diversity order of the MPSK in the hybrid relaying system:
\begin{equation}\label{eq_Diversity_Order}
    G_d^{MPSK} = -\frac{\partial\log(\tilde{P}_{s}^{MPSK})}{\partial\log(\Gamma_1)} = 1.
\end{equation}

\subsection{ASER Performance of DPSK/NCFSK}

\subsubsection{ASER Analysis}
We now analyze the ASER performance of DPSK/NCFSK in the fixed-gain relaying system. For subcarrier DPSK/NCFSK, the conditional SER can be evaluated as \cite{Simon1995}
\begin{equation}\label{eq_ASER_DPSK/NCFSK}
    P_s(e|\gamma) = \frac{1}{2}\exp\left(-\frac{\gamma}{m}\right),
\end{equation}
where $m=1$ for DPSK and $m=2$ for NCFSK. Accordingly, the ASER of DPSK/NCFSK is calculated as
\begin{align}\label{eq_ASER_DPSK}
    P_s^{m} &=\int_0^{\infty}P_s(e|\gamma)f_{\gamma}(\gamma){d}\gamma \overset{(f)}{=} \frac{1}{2m}\int_{0}^{\infty}\exp\left(-\frac{\gamma}{m}\right) F_{\gamma}(\gamma){d}\gamma \nonumber \\
    &=  \frac{1}{2m}\int_0^{\infty} \exp\left(-\frac{\gamma}{m}\right) d{\gamma} - \frac{2^{\alpha}A}{16m\pi} \sum_{k=1}^{\beta}a_k2^{k}\Lambda^{-\frac{\alpha+k}{2}}  \int_{0}^{\infty} \exp(-\frac{\gamma}{m}-\frac{\gamma}{\Gamma_1}) G_{0,5}^{5,0}\left(\frac{\Lambda^2 C\gamma}{16 \Gamma_1\Gamma_2} |_{\kappa}^{-}\right){d}\gamma \nonumber \\
    &\overset{(g)}{=} \frac{1}{2} - \frac{2^{\alpha}A\Gamma_1 }{16\pi(m+\Gamma_1)}\sum_{k=1}^{\beta}a_k2^{k}\Lambda^{-\frac{\alpha+k}{2}} G_{1,5}^{5,1}\left(\frac{m\Lambda^2 C}{16\Gamma_2(m+\Gamma_1)} |_{\kappa_2}^{0}\right),
\end{align}
where Equations $(f)$ and $(g)$ are obtained upon using integration by parts and by exploiting the integral \cite[Eq. (7.813-1)]{Integral_Table}, respectively.

As a special case, for the K-distribution, the ASER of DPSK/NCFSK is given by setting $\rho = 0$ and $\Omega = 0$ in \eqref{eq_ASER_DPSK}, yielding:
\begin{align}\label{eq_ASER_DPSK_K}
   P_s^{m,K} &= \frac{1}{2} - \frac{2^{\alpha}\Gamma_1}{4\pi\Gamma(\alpha)(m+\Gamma_1)} G_{1,5}^{6,1}\left(\frac{m\alpha^2 C}{64 b_0^2 \Gamma_2(m+\Gamma_1)} |_{\kappa_{K}}^{0}\right).
\end{align}
By contrast, for the G-G distribution, we set $\rho = 1$ and $\Omega' = 1$ in \eqref{eq_ASER_DPSK}, yielding:
\begin{align}\label{eq_ASER_DPSK_GG}
    P_s^{m,GG} &= \frac{1}{2} - \frac{2^{\alpha+\beta}\Gamma_1}{8\pi\Gamma(\alpha)\Gamma(\beta)(m+\Gamma_1)} G_{1,5}^{6,1}\left(\frac{m\Lambda^2 C}{16 \Gamma_2(m+\Gamma_1)} |_{\kappa_{GG}}^{0}\right).
\end{align}

\subsubsection{Asymptotic Characterization}
Let us now characterize the asymptotic behavior of DPSK/NCFSK at high SNR. By using the expression in \eqref{eq_MG_HighSNR} and carrying out analogous manipulations to those in \eqref{eq_MGF_HighSNR}, the asymptotic ASER of DPSK/NCFSK at high SNRs is obtained as follows:
\begin{align}\label{eq_ASER_DPSK_HighSNR}
    \tilde{P}_s^{m}&=\lim_{\Gamma_1, \Gamma_2\rightarrow\infty} P_s^{m} =\frac{1}{2} - \frac{A\Gamma_1 }{4(m+\Gamma_1)}\sum_{k=1}^{\beta}a_k\Lambda^{-\frac{\alpha+k}{2}} \Gamma(\alpha)\Gamma(k)\nonumber \\
    &= \frac{1}{2} - \frac{\Gamma_1}{2(m+\Gamma_1)} = \frac{m}{2\Gamma_1} + O\left(\Gamma_1^{-2}\right).
\end{align}

\emph{Remark 2:} From \eqref{eq_ASER_DPSK_HighSNR}, we find that the asymptotic ASER trends of DPSK/NCFSK are similar to those of MPSK. By using the same method, we obtain the diversity order of DPSK/NCFSK as $G_d^{DPSK/NCFSK} = 1$, which is identical to that of MPSK. Now, we can readily compare the performance of different modulation schemes. We define the SNR discrepancy as the difference of the energy per bit to noise ratio between two modulation schemes at a specific ASER. From \eqref{eq_SER_MPSK_Upper_Bound} and \eqref{eq_ASER_DPSK_HighSNR}, the SNR difference between MPSK and DPSK/NCFSK is calculated as follows:
\begin{align}\label{eq_SNR_Gain}
    SNR_{\Delta} &= 10\log_{10}\left(\frac{\Xi}{\tilde{P}_{s}^{MPSK}}\right)- 10\log_{10}\left(\frac{m}{2\tilde{P}_s^{m}}\right) =  10\log_{10}\left(\frac{m}{2\Xi}\right)~ \textrm{dB}.
\end{align}
For example, we calculate the $SNR_{\Delta}$ of BPSK with respect to DPSK by setting $M = 2$ and $m = 1$ in \eqref{eq_SNR_Gain}. The SNR gain of BPSK over DPSK becomes
\begin{equation}\label{eq_SNR_Gain_QPSK_DPSK}
    SNR_{\Delta}^{BPSK-DPSK} = 10\log_{10}\left(\frac{24}{13}\right) = 2.66 ~ \textrm{dB}.
\end{equation}
It is observed from the above equation that, as expected, BPSK achieves a better SER performance than DPSK, because the phase error in DPSK induces an error in two consecutive transmission intervals. Similarly, the $SNR_{\Delta}$ between different PSK schemes can be evaluated by using \eqref{eq_SNR_Gain}. For instance, we can readily express the SNR reduction of BPSK over QPSK as $SNR_{\Delta}^{BPSK-QPSK} = 10\log_{10}\left(\frac{44}{13}\right) = 5.3 ~ \textrm{dB}$ at a specific ASER, albeit naturally, QPSK achieves a factor of two higher bandwidth efficiency in the fixed-gain relaying system.

\section{Performance of the Hybrid RF/FSO System with Channel-Dependent Relaying}
In some applications, the relaying system has a strict by limited transmit power for forwarding signals. This implies that the maximum energy of the signal forwarded by the RN should be limited. In such applications, the RN dynamically adjusts the relay gain according to the instantaneous channel fading of the RF link. In this section, the ASERs of various modulation schemes are investigated analytically for the channel-dependent hybrid relaying system.

\subsection{ASER Performance of MPSK}
To analyze the ASER performance of MPSK in the channel-dependent relaying system context, we firstly derive some useful statistical features of the end-to-end SNR. Since the closed-form analytical expressions of the SNR statistics are intractable, we resort to the derivation of tight upper bounds of the SNR statistics.

\subsubsection{Statistical Characteristics}
The end-to-end SNR $\gamma$ in the hybrid channel-dependent relaying system may be closely approximated as \cite{Peppas2013, Zedini2015}:
\begin{equation}\label{eq_SNR_CSI}
    \gamma = \frac{\gamma_1\gamma_2}{\gamma_1+\gamma_2+1} \overset{\sim}{=} \min\{\gamma_1,\gamma_2\}.
\end{equation}
Accordingly, the CDF of the end-to-end SNR $\gamma$ in the channel-dependent relaying system can be expressed as:
\begin{align}\label{eq_CDF_CSI}
    F_{\gamma}(\gamma)& = Pr[\min\{\gamma_1,\gamma_2\} \leq \gamma] \nonumber \\
    &= 1 -  Pr(\gamma_1 > \gamma) Pr(\gamma_2 > \gamma)\nonumber \\
    &=  F_{\gamma_1}(\gamma) + F_{\gamma_2}(\gamma) - F_{\gamma_1}(\gamma) F_{\gamma_2}(\gamma),
\end{align}
where $F_{\gamma_1}(\gamma)$ and $F_{\gamma_2}(\gamma)$ denote the CDFs of the RF link and of the FSO link, respectively. For the RF link, the CDF of the SNR $\gamma_1$ can be readily calculated from \eqref{eq_Rayleigh_PDF} as follows:
\begin{align}\label{eq_CDF_Rayleigh_CSI}
    F_{\gamma_1}(\gamma) &= \frac{1}{\Gamma_1} \int_{0}^{\gamma}\exp\left(-\frac{\gamma_1}{\Gamma_1}\right){d}\gamma_1 = 1-\exp\left(-\frac{\gamma}{\Gamma_1}\right).
\end{align}
Meanwhile, by using \eqref{eq_M_SNR_PDF}, the CDF of the SNR $\gamma_2$ for the FSO link is calculated as follows:
\begin{align}\label{eq_CDF_M_CSI}
  F_{\gamma_2}(\gamma) &= \frac{A}{2}\sum_{k=1}^\beta a_k \Gamma_2^{-\frac{\alpha+k}{4}} \int_{0}^{\gamma} \gamma_2^{\frac{\alpha+k}{4}-1} K_{\alpha-k}\left(2\sqrt{\Lambda\sqrt{\frac{\gamma_2}{\Gamma_2}}}\right){d}\gamma_2 \nonumber \\
  &\overset{(h)}{=} \frac{2^{\alpha}A}{8\pi}\sum_{k=1}^\beta a_k 2^k \Lambda^{-\frac{\alpha+k}{2}} G_{1,5}^{4,1 }\left(\frac{\Lambda^2\gamma}{16\Gamma_2}|_{\kappa_2}^{1}\right),
\end{align}
where Equation $(h)$ is calculated using \cite[Eq. 07.34.21.0084.01, Eq. 07.34.17.0011.01]{Wolfram}. Upon substituting \eqref{eq_CDF_Rayleigh_CSI} and \eqref{eq_CDF_M_CSI} into \eqref{eq_CDF_CSI}, the CDF of the end-to-end SNR $\gamma$ for the channel-dependent relaying system is arrived at:
\begin{align}\label{eq_CDF_CSI_Re}
    F_{\gamma}(\gamma)&= 1 - \exp\left(-\frac{\gamma}{\Gamma_1}\right) + \frac{2^{\alpha}A}{8\pi}\exp\left(-\frac{\gamma}{\Gamma_1}\right) \sum_{k=1}^\beta a_k 2^k \Lambda^{-\frac{\alpha+k}{2}} G_{1,5}^{4,1 }\left(\frac{\Lambda^2\gamma}{16\Gamma_2}|_{\kappa_2}^{1}\right).
\end{align}

From \eqref{eq_CDF_CSI_Re}, the MGF of the end-to-end SNR $\gamma$ of the channel-dependent relaying system can be derived as:
\begin{align}\label{eq_MGF_CSI}
    M_{\gamma}(s) &=  s\int_{0}^{\infty}e^{-s\gamma}F_{\gamma}(\gamma){d}\gamma \nonumber \\
    &= s \int_0^{\infty} \exp(-\gamma s)d{\gamma} - s \int_0^{\infty} \exp(-\gamma s-\frac{\gamma}{\Gamma_1})d{\gamma} \nonumber \\
    & \quad + \frac{2^{\alpha}As}{8\pi}\sum_{k=1}^{\beta}a_k 2^k \Lambda^{-\frac{\alpha+k}{2}} \int_{0}^{\infty}  \exp(-s\gamma-\frac{\gamma}{\Gamma_1})G_{1,5}^{4,1 }\left(\frac{\Lambda^2\gamma}{16\Gamma_2}|_{\kappa_2}^{1}\right) {d}\gamma \nonumber \\
    & \overset{(i)}{=} 1 - \frac{\Gamma_1s}{1+\Gamma_1s} + \frac{2^{\alpha}A\Gamma_1s}{8\pi (1+\Gamma_1s)}\sum_{k=1}^{\beta}a_k2^{k}\Lambda^{-\frac{\alpha+k}{2}}  G_{2,5}^{4,2}\left(\frac{\Lambda^2 \Gamma_1}{16 \Gamma_2(1+\Gamma_1s)} |_{\kappa_2}^{0, 1}\right) \nonumber \\
    & \overset{(j)}{=} 1 - \frac{\Gamma_1s}{1+\Gamma_1s} + \frac{2^{\alpha}A\Gamma_1s}{8\pi (1+\Gamma_1s)}\sum_{k=1}^{\beta}a_k2^{k}\Lambda^{-\frac{\alpha+k}{2}}  G_{1,4}^{4,1}\left(\frac{\Lambda^2 \Gamma_1}{16 \Gamma_2(1+\Gamma_1s)} |_{\kappa_3}^{1}\right),
\end{align}
where we have $\kappa_3 = \frac{\alpha}{2},\frac{\alpha+1}{2}, \frac{k}{2}, \frac{k+1}{2}$, and Equations (i) as well as (j) are obtained using \cite[Eq. 07.34.21.0088.01]{Wolfram} and \cite[Eq. 9.31-1]{Integral_Table}, respectively.

\subsubsection{ASER Analysis}
In the hybrid channel-dependent relaying system, the ASER of MPSK can be evaluated by substituting \eqref{eq_MGF_CSI} into \eqref{eq_SEP_MPSK}, which yields:
\begin{align}\label{eq_SER_MPSK_CSI}
    P_s^{MPSK} &= \frac{1}{\pi}\int_0^{\Theta}1 {d}\theta - \frac{\Gamma_1n^2}{\pi}\int_0^{\Theta}\frac{1}{\Phi} {d}\theta + \frac{2^{\alpha}A\Gamma_1n^2}{8\pi^2} \sum_{k=1}^{\beta}a_k2^{k} \Lambda^{-\frac{\alpha+k}{2}} \int_0^{\Theta}\frac{1}{\Phi} G_{1,4}^{4,1}\left(\frac{\Lambda^2 \Gamma_1 \sin^2\theta}{16 \Gamma_2\Phi} |_{\kappa_3}^{1}\right){d}\theta.
\end{align}

The integral above cannot be readily expressed in a closed form. Using a technique analogous to that of the previous section, we can instead calculate the approximate ASER expression of MPSK by substituting \eqref{eq_MGF_CSI} into \eqref{eq_SER_MPSK_Approx} for the hybrid channel-dependent relaying system.

As stated above, both the K and G-G distributions can be deduced from the generalized M-distribution. To derive the ASER of MPSK for those channels, we obtain the MGF of the end-to-end SNR $\gamma$ over the K and G-G-distributed FSO channels from \eqref{eq_MGF_CSI} as follows:
\begin{align}\label{eq_MGF_CSI_K}
    M_{\gamma}^K(s)  &= 1 - \frac{\Gamma_1s}{1+\Gamma_1s} + \frac{2^{\alpha}\Gamma_1s}{2\pi\Gamma(\alpha) (1+\Gamma_1s)} G_{1,4}^{4,1}\left(\frac{\Lambda^2 \Gamma_1}{16 \Gamma_2(1+\Gamma_1s)} |_{\kappa_{3K}}^{1}\right),
\end{align}

\begin{align}\label{eq_MGF_CSI_GG}
    M_{\gamma}^{GG}(s) &= 1 - \frac{\Gamma_1s}{1+\Gamma_1s} + \frac{2^{\alpha+\beta}\Gamma_1s}{4\pi\Gamma(\alpha)\Gamma(\beta)(1+\Gamma_1s)} G_{1,4}^{4,1}\left(\frac{\Lambda^2 \Gamma_1}{16 \Gamma_2(1+\Gamma_1s)} |_{\kappa_{3GG}}^{1}\right),
\end{align}
where we have $\kappa_{3K} = \frac{\alpha}{2}, \frac{\alpha+1}{2}, \frac{1}{2}, 1 $ and $\kappa_{3GG} = \frac{\alpha}{2},\frac{\alpha+1}{2}, \frac{\beta}{2}, \frac{\beta+1}{2}$. Accordingly, both the exact and the approximate ASER of MPSK can be derived for transmission over the K and G-G channels in the channel-dependent relaying system by substituting \eqref{eq_MGF_CSI_K} and \eqref{eq_MGF_CSI_GG} into \eqref{eq_SEP_MPSK} and \eqref{eq_SER_MPSK_Approx}, respectively.

\subsubsection{Asymptotic Characterization}
Similar to the fixed-gain relaying system, we may resort to characterizing the asymptotic approximation of the MGF at high SNRs, which proves beneficial for obtaining the asymptotic ASER for further insightful observations. By using the series representation of Meijer's G-function for $\Gamma_1, \Gamma_2 \rightarrow \infty$ and considering the fact that $\alpha$ is larger than $\beta$ in the M-distribution, the series expansion of Meijer's G-function is dominated by the smallest parameter in $\kappa_3$, which yields:
\begin{align}\label{eq_MG_CSI_HighSNR}
& \quad\lim_{\Gamma_1, \Gamma_2\rightarrow\infty} G_{1,4}^{4,1}\left(\frac{\Lambda^2 \Gamma_1}{16 \Gamma_2 (1+\Gamma_1 s)} |_{\kappa_3}^{1}\right) \nonumber \\
    & = \sqrt{\pi}\Gamma\left(\frac{\alpha-k}{2}\right) \Gamma\left(\frac{\alpha-k+1}{2}\right) \Gamma\left(\frac{k}{2}\right)  \left(\frac{\Lambda^2 \Gamma_1}{16 \Gamma_2(1+\Gamma_1 s)}\right)^{\frac{k}{2}} \nonumber \\% + O\left(\frac{\Lambda^2 C_2}{16 \Gamma_2(1+\Gamma_1s)}\right)
    & = 2^{1-\alpha+k}\pi\Gamma(\alpha-k)\Gamma\left(\frac{k}{2}\right)\left(\frac{\Lambda^2 \Gamma_1}{16 \Gamma_2(1+\Gamma_1 s)}\right)^{\frac{k}{2}},
\end{align}
where the last equation is obtained by exploiting $\Gamma(2x) = \frac{2^{2x-1}}{\sqrt{\pi}}\Gamma(x)\Gamma(x+1/2)$. By substituting \eqref{eq_MG_CSI_HighSNR} into \eqref{eq_MGF_CSI}, the asymptotic MGF becomes:
\begin{align}\label{eq_MGF_Asymp_CSI}
    \tilde{M}_{\gamma}(s) &= \lim_{\Gamma_1, \Gamma_2\rightarrow \infty} M_{\gamma}(s) = \frac{1}{1+\Gamma_1 s} + \frac{A \Gamma_1 s}{4(1+\Gamma_1 s)} \sum_{k=1}^{\beta} a_k \Lambda^{\frac{k-\alpha}{2}}\Gamma(\alpha-k)  \Gamma\left(\frac{k}{2}\right)\left(\frac{\Gamma_1}{\Gamma_2(1+\Gamma_1 s)}\right)^{\frac{k}{2}} \nonumber \\
    &= \sum_{k=1}^{\beta} B_k \left(\Gamma_2 s\right)^{-\frac{k}{2}},
\end{align}
where we have $B_k = \frac{A}{4} a_k \Lambda^{\frac{k-\alpha}{2}}\Gamma(\frac{k}{2})\Gamma(\alpha-k)$.

Now, by substituting \eqref{eq_MGF_Asymp_CSI} into \eqref{eq_SER_MPSK_Approx}, we obtain the asymptotic ASER of MPSK in the channel-dependent relaying system context as:
\begin{align}\label{eq_SER_MPSK_Approx_CSI}
    \tilde{P_s}^{MPSK} &= \left(\frac{M-1}{2M}-\frac{1}{6}\right)\sum_{k=1}^{\beta} B_k \left(\Gamma_2 s_1\right)^{-\frac{k}{2}} +\frac{1}{4}\sum_{k=1}^{\beta} B_k \left(\Gamma_2 s_2\right)^{-\frac{k}{2}}+ \left(\frac{M-1}{2M}-\frac{1}{4}\right) \sum_{k=1}^{\beta} B_k \left(\Gamma_2 s_3\right)^{-\frac{k}{2}} \nonumber \\
    &= B_1 D_M\Gamma_2^{-\frac{1}{2}},
\end{align}
%where $D_M = \left[ \left(\frac{M-1}{2M}-\frac{1}{6}\right) s_1^{-\frac{1}{2}}+\frac{1}{4}s_2^{-\frac{1}{2}} + \left(\frac{M-1}{2M}-\frac{1}{4}\right) s_3^{-\frac{1}{2}}\right]$
where $D_M = \frac{8M+3\sqrt{3}M+(6M-12)\sqrt{s_1} - 12}{24M\sqrt{s_1}}$ is a constant dependent on the modulation order, while the short-form equation in the 2nd line is obtained by retaining the most dominant term for high SNRs.

\emph{Remark 3:} It is observed from \eqref{eq_SER_MPSK_Approx_CSI} that the ASER of MPSK depends both on the average SNR and on the AT conditions of the FSO link, while the diversity order of MPSK in the channel-dependent relaying system equals to:
 \begin{equation}\label{eq_Diversity_Order_CSI}
    G_d^{MPSK} =  -\frac{\partial\log(\tilde{P}_{s}^{MPSK})}{\partial\log(\Gamma_2)} = \frac{1}{2}.
 \end{equation}
However, the diversity order in \eqref{eq_Diversity_Order_CSI} does not hold under the G-G distribution ($\rho = 1$), since the asymptotic expression of the Meijer's-G function in \eqref{eq_MG_CSI_HighSNR} does not converge for $\beta > 1$ at high SNRs.

\subsection{ASER of DPSK/NCFSK}

\subsubsection{ASER Analysis}
Using \eqref{eq_ASER_DPSK/NCFSK}, the ASER of DPSK/NCFSK in the channel-dependent relaying system can be derived as
\begin{align}\label{eq_ASER_DPSK_CSI}
    P_s^{m} &= \frac{1}{2m}\int_{0}^{\infty}\exp\left(-\frac{\gamma}{m}\right) F_{\gamma}(\gamma){d}\gamma \nonumber \\
    %&=  \frac{1}{2m}\int_0^{\infty} \exp\left(-\frac{\gamma}{m}\right) d{\gamma} - \frac{1}{2m}\int_0^{\infty} \exp\left(-\frac{\gamma}{m}-\frac{\gamma}{\Gamma_1}\right) d{\gamma} \nonumber \\
%    & \quad + \frac{2^{\alpha}A}{16m\pi} \sum_{k=1}^{\beta}a_k2^{k}\Lambda^{-\frac{\alpha+k}{2}} \nonumber \\
%    & \quad \times  \int_{0}^{\infty} \exp(-\frac{\gamma}{m}-\frac{\gamma}{\Gamma_1}) G_{1,5}^{4,1}\left(\frac{\Lambda^2 \gamma}{16\Gamma_2} |_{\kappa_2}^{1}\right){d}\gamma \nonumber \\
    &= \frac{1}{2} - \frac{\Gamma_1}{2(m+\Gamma_1)} + \frac{2^{\alpha}A\Gamma_1 }{16\pi(m+\Gamma_1)}\sum_{k=1}^{\beta}a_k2^{k}\Lambda^{-\frac{\alpha+k}{2}}  G_{1,4}^{4,1}\left(\frac{\Lambda^2\Gamma_1 m}{16\Gamma_2(m+\Gamma_1)} |_{\kappa_3}^{1}\right),
\end{align}
where the last equation is calculated using \cite[Eq. 07.34.21.0088.01]{Wolfram} and \cite[Eq. 9.31-1]{Integral_Table}.

Specifically, for the K and G-G distributed FSO channels, the ASER of DPSK/NCFSK in the hybrid channel-dependent relaying system is readily calculated as
\begin{align}\label{eq_ASER_DPSK_CSI_K}
    P_s^{m,K} &= \frac{1}{2} - \frac{\Gamma_1}{2(m+\Gamma_1)} + \frac{2^{\alpha}\Gamma_1 }{4\pi\Gamma(\alpha)(m+\Gamma_1)} G_{1,4}^{4,1}\left(\frac{\Lambda^2\Gamma_1 m}{16\Gamma_2(m+\Gamma_1)} |_{\kappa_{3K}}^{1}\right),
\end{align}
and as
\begin{align}\label{eq_ASER_DPSK_CSI_GG}
    P_s^{m,GG} &= \frac{1}{2} - \frac{\Gamma_1}{2(m+\Gamma_1)} + \frac{2^{\alpha+\beta}\Gamma_1 }{8\pi\Gamma(\alpha)\Gamma(\beta)(m+\Gamma_1)} G_{1,4}^{4,1}\left(\frac{\Lambda^2\Gamma_1 m}{16\Gamma_2(m+\Gamma_1)} |_{\kappa_{3GG}}^{1}\right),
\end{align}
respectively.

\subsubsection{Asymptotic Characterization}
We analyze the asymptotic ASER of DPSK/NCFSK in the channel-dependent relaying system. By using the analogous expression in \eqref{eq_MG_CSI_HighSNR}, and following some further manipulations, the asymptotic ASER of DPSK/NCFSK at high SNRs can be expressed as
\begin{align}\label{eq_SER_DPSK_Approx_CSI}
    \tilde{P_s}^{m} &= \lim_{\Gamma_1, \Gamma_2 \rightarrow \infty} P_s^m = \frac{1}{2(m+\Gamma_1)} + \frac{A \Gamma_1}{8(m+\Gamma_1)} \sum_{k=1}^{\beta} a_k \Lambda^{\frac{k-\alpha}{2}}\Gamma(\alpha-k)  \Gamma\left(\frac{k}{2}\right)\left(\frac{m\Gamma_1}{\Gamma_2(m+\Gamma_1)}\right)^{\frac{k}{2}} \nonumber \\
    &= \frac{1}{2} \sum_{k=1}^{\beta} B_k \left(\frac{m}{\Gamma_2}\right)^{\frac{k}{2}} = \frac{B_1\sqrt{m}}{2} \Gamma_2^{-\frac{1}{2}},
\end{align}
where again, the last short-form equation is obtained by retaining the most dominant term at a high SNR. It can be seen from the above expression that the diversity order of DPSK/NCFSK is identical to that of MPSK in the channel-dependent relaying system, and that the ASER of DPSK/NCFSK depends both on the average SNR and on the channel conditions of the FSO link.

\emph{Remark 4:} We can compare the ASER performance of different modulation schemes in the context of the channel-dependent relaying system. Similar to \eqref{eq_SNR_Gain}, we can readily calculate the SNR difference between MPSK and DPSK/NCFSK as follows:
\begin{align}\label{eq_SNR_Gain_CSI}
  SNR_{\Delta} &= 10\log_{10}\left(\frac{B_1D_M}{\tilde{P_s}^{MPSK}}\right)^2- 10\log_{10}\left(\frac{B_1 \sqrt{m}}{2\tilde{P_s}^{m}}\right)^2 = 10\log_{10}\left(\frac{4D_M^2}{m}\right) \mathrm{dB}.
\end{align}
For instance, we obtain the $SNR_{\Delta}$ of BPSK with respect to DPSK by setting $M = 2$ and $m = 1$ in \eqref{eq_SNR_Gain_CSI}, yielding $SNR_{\Delta}^{BPSK-DPSK} = 4.4$ dB.

\emph{Remark 5:} It is interesting to observe from \eqref{eq_Diversity_Order} and \eqref{eq_Diversity_Order_CSI} that all the modulation schemes in the fixed-gain relaying system achieve the same diversity order of 1, while the diversity order in the channel-dependent relaying system is only $\frac{1}{2}$. This is because the RN can amplify the signal received from the first link with a high relay gain regardless of the channel condition of the first link, thus it is capable of achieving a higher diversity order at the cost of requiring a high-dynamic linear class-A PA. By contrast, in the channel-dependent relaying strategy, the relay gain is dynamically adjusted according to the CSI of the first link in order to maintain a constant output power at the RN. For instance, if the channel gain of the first link happens to be high, the relay gain at the RN will be much smaller for the channel-dependent strategy in order to guarantee a constant RN output power, which imposes a lower diversity order compared to that of the fixed-gain relay strategy.

\section{Simulation Results}
In this section, we detail our simulation results for the dual-hop hybrid RF/FSO relaying system relying on both the fixed-gain and the channel-dependent schemes. The transmitted optical power is normalized to unity, i.e. we have $\Omega + 2 b_0 = 1$. For fixed-gain relaying scheme, the relaying gain is set such as $C = 0.5$. For simplicity and without loss of generality, we opt for an identical average SNR per hop, i.e. for $\Gamma_1 = \Gamma_2$.

\begin{figure}\centering
  \includegraphics[scale = 0.5]{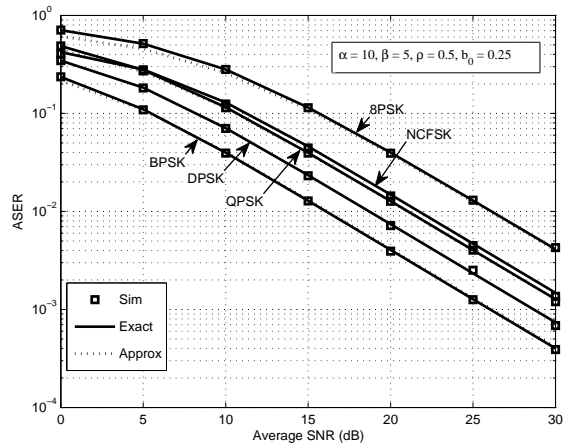}
  \caption{ASER versus average SNR of different modulation schemes for the fixed-gain relaying system. }\label{Fig_Different_Mod_AF}
\end{figure}

The ASER versus average SNR of different modulation schemes for the fixed-gain relaying system is presented in Fig.~\ref{Fig_Different_Mod_AF}, where we have $\alpha = 10, \beta = 5, \rho = 0.5$, and $b_0 = 0.25$. The simulation results (referred to as `Sim') recorded for both MPSK and DPSK/NCFSK were obtained by Monte Carlo simulations, the exact ASER (`Exact') and the approximate ASER (`Approx') of MPSK were calculated from \eqref{eq_SEP_MPSK_Re} and \eqref{eq_SER_MPSK_Approx}, respectively. Finally, the exact ASER of DPSK/NCFSK was calculated from \eqref{eq_ASER_DPSK}. From this figure, we find that the exact ASER perfectly matches the simulation results for both MPSK and DPSK, which confirms the accuracy of our analytical results. Furthermore, the approximate ASER exhibits an excellent agreement with the exact results for different PSK schemes, which confirms the accuracy of our approximate expressions. Moreover, high-order PSK achieves an increased throughput at the expense of a degraded ASER performance compared to low-order PSK. On the other hand, the performance of DPSK is inferior to BPSK, but is superior to NCFSK. It is observed from this figure that the SNR penalty of DPSK compared to BPSK is about 3 dB, again, because the phase error in DPSK induces the bit error in two consecutive transmission intervals, which donates the expected SER of DPSK compared to BPSK at a specific SNR. To the contrary, the SNR gain of DPSK compared to NCFSK is about 3 dB, because the coherent detection can achieve a SNR gain of 3 dB compared to the non-coherent detection at a specific SNR. However, the coherent detection needs channel estimation at the detector, which might be more complexity than the non-coherent detection.

\begin{figure}\centering
  \includegraphics[scale = 0.5]{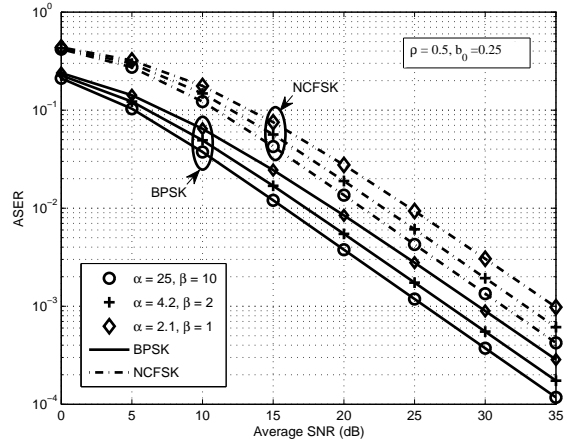}
  \caption{SER versus average SNR for different values of $\alpha, \beta$ in the fixed-gain relaying system. }\label{Fig_Alpha_Beta}
\end{figure}

The ASER versus average SNR is presented in Fig.~\ref{Fig_Alpha_Beta} for different values of $\alpha$ and $\beta$ in the fixed-gain relaying system, where we have $\rho = 0.5$, $b_0 = 0.25$. To avoiding obfuscating legends, the Monte Carlo verification points are not shown in some of the following figures. The ASERs of two PSK schemes were calculated by substituting \eqref{eq_MGF_Re} into \eqref{eq_SER_MPSK_Approx}. As we can see from Fig.~\ref{Fig_Alpha_Beta}, the ASER performance improves upon increasing the AT parameters $\alpha$, $\beta$. This is because the turbulence becomes less severe upon increasing $\alpha$, $\beta$, hence the SER performance of both modulation schemes becomes better.

\begin{figure}\centering
 \includegraphics[scale = 0.5]{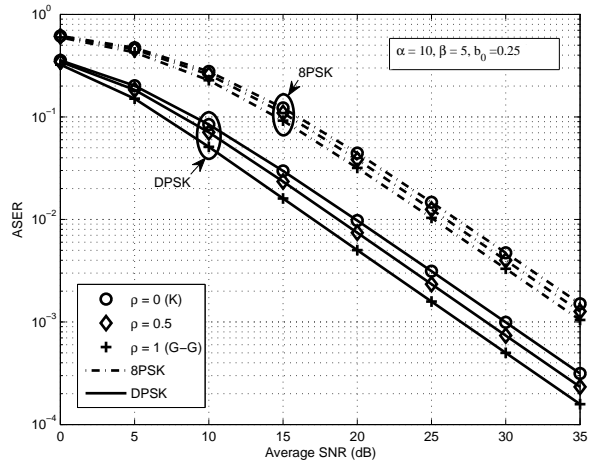}
 \caption{ASER versus average SNR for various $\rho$ values in the fixed-gain relaying system.}\label{Fig_M_Rho}
\end{figure}

The ASER versus average SNR performances of 8PSK and DPSK is shown in Fig.~\ref{Fig_M_Rho} for different $\rho$ values in the fixed-gain relaying system, where we have $\alpha = 10$, $\beta = 5$, and $b_0 = 0.25$. %To avoiding obfuscating legends, the Monte Carlo verification points are not shown in some of the following figures.
The ASER of 8PSK under generalized M as well as K and G-G distributed channels was calculated by substituting \eqref{eq_MGF_Re}, \eqref{eq_MGF_K} and \eqref{eq_MGF_GG} into \eqref{eq_SER_MPSK_Approx}, respectively. The ASER of DPSK under generalized M as well as K and G-G channels was calculated from \eqref{eq_ASER_DPSK}, \eqref{eq_ASER_DPSK_K} and \eqref{eq_ASER_DPSK_GG}, respectively. It is observed that the ASER is reduced for both modulation schemes upon decreasing $\rho$, because the average power of the classic scattering component equals $\xi = 2b_0(1-\rho)$ and it becomes higher as $\rho$ is reduced. Increasing the classic scattering component power will make the intensity fluctuations of the optical signal more severe along the transmission path. Specifically, when we have $\rho = 1$ and $\Omega' = 1$, the generalized M-distribution reduces to the G-G distribution. In this situation, there is no classic scattering component, because only the LoS component and the coupled scattering component exists. By contrast, the classic scattering component reaches its maximal value at $\rho = 0$ and it degrades the ASER performance. In this case, the generalized M-distribution reduces to the K-distribution.

\begin{figure}\centering
  \includegraphics[scale = 0.5]{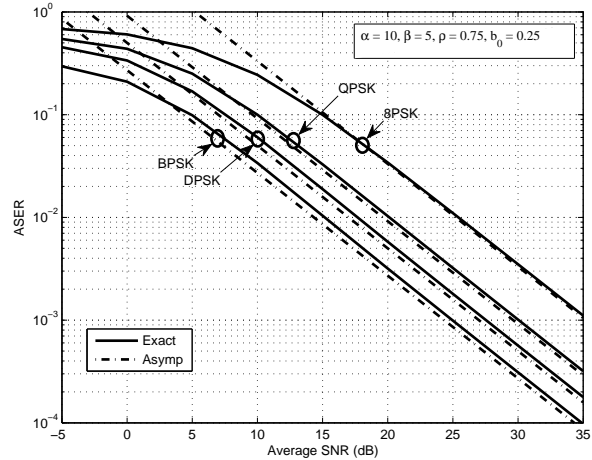}
  \caption{Asymptotic ASER versus average SNR of different modulation schemes for the fixed-gain relaying system. }\label{Fig_ASER_Asymp}
\end{figure}

Let us now study the asymptotic ASER (`Asymp') of different modulation schemes for the fixed-gain relaying system in Fig.~\ref{Fig_ASER_Asymp}, where we have $\alpha = 10$, $\beta = 5$, $\rho = 0.75$, and $b_0 = 0.25$. The asymptotic ASERs of MPSK and DPSK were calculated from \eqref{eq_SER_MPSK_Upper_Bound} and \eqref{eq_ASER_DPSK_HighSNR}, respectively, while the exact ASERs of MPSK and DPSK were calculated from \eqref{eq_SEP_MPSK_Re} and \eqref{eq_ASER_DPSK}, respectively. It is observed in Fig.~\ref{Fig_ASER_Asymp} that the asymptotic ASERs are consistent with the exact ASERs for both MPSK and DPSK at high SNRs. As seen in Fig.~\ref{Fig_ASER_Asymp}, there is only a modest gap between the asymptotic and the exact ASER of the different modulation schemes. The asymptotic expressions of \eqref{eq_SER_MPSK_Upper_Bound} and \eqref{eq_ASER_DPSK_HighSNR} are much simpler, which also reveals that both MPSK and DPSK/NCFSK have the same diversity order. Furthermore, it is intuitive that the high order modulation schemes require a higher SNR in order to maintain a specific ASER. For instance, the SNR losses of 8PSK compared to QPSK and BPSK are 5.7 dB and 11 dB, respectively. which is the price of increasing the throughput from 1 to 3 bits/symbol.

%\begin{figure}\centering
%  \includegraphics[scale = 0.5]{Asymmetric_SNR.eps}
%  \caption{ ASER versus average SNR of different modulation schemes with asymmetric SNR at both hops in the fixed-gain relaying system. }\label{Fig_Asymmetic_SNR}
%\end{figure}
%
%Let us now study the asymptotic ASER (`Asymp') of different modulation schemes for the

\begin{figure}\centering
  \includegraphics[scale = 0.5]{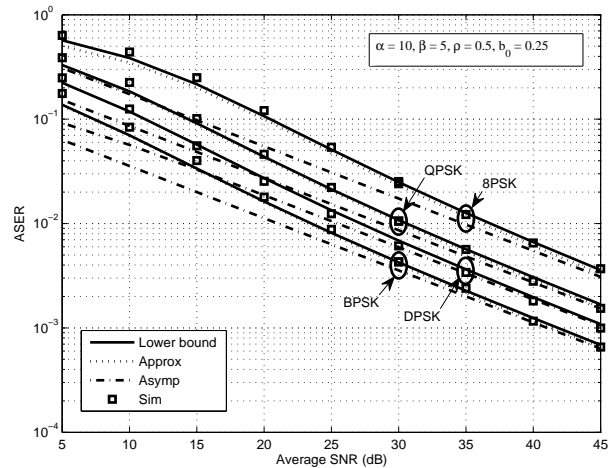}
  \caption{ASER versus average SNR of different modulation schemes for the channel-dependent relaying system. }\label{Fig_Different_Mod_CSI}
\end{figure}

We illustrate the ASER performance of different modulation schemes for the channel-dependent relaying system in Fig.~\ref{Fig_Different_Mod_CSI}, where we have $\alpha = 10$, $\beta = 5$, $\rho = 0.5$, and $b_0 = 0.25$. The simulation results recorded for both MPSK and DPSK/NCFSK were obtained by Monte Carlo simulations, while the lower bound ASERs (`Lower bound') of MPSK and DPSK/NCFSK were calculated from \eqref{eq_SER_MPSK_CSI} and \eqref{eq_ASER_DPSK_CSI}, respectively. The approximate ASER of MPSK was obtained by substituting \eqref{eq_MGF_CSI} into \eqref{eq_SER_MPSK_Approx}. Finally, the asymptotic ASERs of MPSK and DPSK/NCFSK were calculated from \eqref{eq_SER_MPSK_Approx_CSI} and \eqref{eq_SER_DPSK_Approx_CSI}, respectively. It is observed from this figure that there is a modest gap between the lower bounded ASER and the simulation results for all modulation schemes at lower SNRs, whereas the lower bound ASER moves close to the simulation results, as the SNR increases. The approximate ASER exhibits an excellent agreement with the exact results for the different PSK schemes, which confirms the accuracy of our approximate expressions. Furthermore, the asymptotic results of all modulation schemes approach the exact results upon increasing the SNR.

\begin{figure}\centering
  \includegraphics[scale = 0.5]{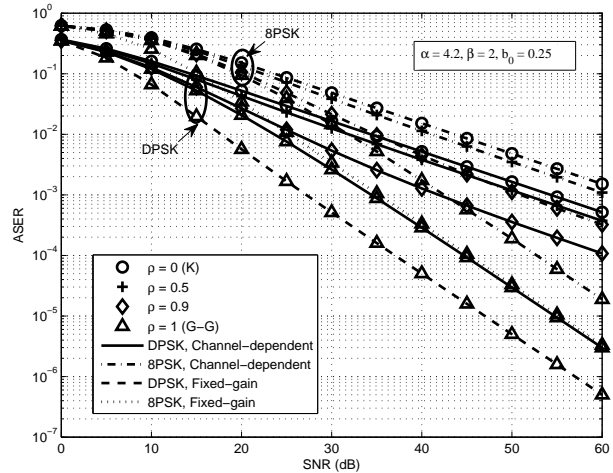}
  \caption{ASER versus average SNR for different $\rho$ values.}\label{Fig_Different_rho_CSI}
\end{figure}

The ASER performance of 8PSK and DPSK is portrayed in Fig.~\ref{Fig_Different_rho_CSI} for different $\rho$ values, where we have $\alpha = 4.2$, $\beta = 2$, and $b_0 = 0.25$.  For comparison, we also draw the ASER performance of DPSK and 8PSK in the fixed-gain relaying system at $\rho = 1$ by using \eqref{eq_ASER_DPSK_GG} and by substituting \eqref{eq_MGF_GG} into \eqref{eq_SER_MPSK_Approx}, respectively. For the channel-dependent relaying system, the ASER of 8PSK under generalized M- as well as K- and G-G distributed channels was calculated by substituting \eqref{eq_MGF_CSI}, \eqref{eq_MGF_CSI_K} and \eqref{eq_MGF_CSI_GG} into \eqref{eq_SER_MPSK_Approx}, respectively. The ASER of DPSK under generalized M as well as K and G-G channels in the channel-dependent relaying system was calculated from \eqref{eq_ASER_DPSK_CSI}, \eqref{eq_ASER_DPSK_CSI_K} and \eqref{eq_ASER_DPSK_CSI_GG}, respectively.

We can find from Fig.~\ref{Fig_Different_rho_CSI} that the ASER improves for both modulation schemes upon increasing $\rho$. This is because the average power of the classic scattering component decreases as $\rho$ increases. For example, both modulation schemes achieve the best ASER performance under the G-G distribution, i.e. $\rho = 1$. This is due to the fact that no classic scattering component exists in the G-G channel. In this situation, both modulation schemes of the channel-dependent relaying system achieve the same diversity order as that in the fixed-gain relaying system, as shown in Fig.~\ref{Fig_Different_rho_CSI}. By contrast, the diversity orders under other situations, such as $\rho = 0$, $0.5$, and $0.9$, are identical to $1/2$ as shown in \eqref{eq_Diversity_Order_CSI} and \eqref{eq_SER_DPSK_Approx_CSI}. Furthermore, it is observed from Fig.~\ref{Fig_Different_rho_CSI} that the ASER performance of the fixed-gain relaying system is superior to that of the channel-dependent relaying system under the G-G distribution. This can be explained as follows: In the channel-dependent relaying system, the relay gain is dynamically adjusted according to the CSI of the first link in order to maintain a constant output power at the relay node (RN). By contrast, in the fixed-gain relaying system, the RN amplifies the signal received from the first link with a high relay gain regardless of the channel condition of the first link. Therefore, the fixed-gain relaying system always outperforms the channel-dependent relaying system at the cost of requiring a high-dynamic linear class-A power amplifier.

\section{Conclusions}\label{section_conclusion}
This paper investigated the ASER performance of different modulation schemes in a dual-hop hybrid RF/FSO relaying system operated under both a fixed-gain and a channel-dependent scheme. The RF and the FSO links were subjected to Rayleigh- and M-distributed impairments, respectively, and we then derived the MGF of the end-to-end instantaneous SNR. Accordingly, both the ASER expressions of MPSK and DPSK/NCFSK were obtained. Furthermore, the ASER experienced in the presence of both K and G-G distributions were also evaluated as special cases of our results. Finally, the asymptotic ASER of both MPSK and DPSK/NCFSK were also derived at high SNRs, which demonstrated some insightful observations for both hybrid relaying systems. Our simulation results confirm the accuracy of our analytical results.

\bibliographystyle{IEEEtran}
\bibliography{IEEEabrv,Reference}

\end{document}